\documentclass[letterpaper,prl,twocolumn,superscriptaddress]{revtex4}
\usepackage{graphicx}
\usepackage{times}
\begin{document}
\title{Squeezed light at sideband frequencies below 100 kHz from a single OPA}

\author{R.  Schnabel}
\affiliation{Max-Planck-Institut f\"ur Gravitationsphysik (Albert-Einstein-Institut),\\
Institut f\"ur Atom- und Molek\"ulphysik, Universit\"at Hannover, 30167 Hannover, Germany}

\author{H. Vahlbruch}
\affiliation{Max-Planck-Institut f\"ur Gravitationsphysik (Albert-Einstein-Institut),\\
Institut f\"ur Atom- und Molek\"ulphysik, Universit\"at Hannover, 30167 Hannover, Germany}

\author{A. Franzen}
\affiliation{Max-Planck-Institut f\"ur Gravitationsphysik (Albert-Einstein-Institut),\\
Institut f\"ur Atom- und Molek\"ulphysik, Universit\"at Hannover, 30167 Hannover, Germany}

\author{S. Chelkowski}
\affiliation{Max-Planck-Institut f\"ur Gravitationsphysik (Albert-Einstein-Institut),\\
Institut f\"ur Atom- und Molek\"ulphysik, Universit\"at Hannover, 30167 Hannover, Germany}

\author{N. Grosse}
\affiliation{Max-Planck-Institut f\"ur Gravitationsphysik (Albert-Einstein-Institut),\\
Institut f\"ur Atom- und Molek\"ulphysik, Universit\"at Hannover, 30167 Hannover, Germany}

\author{H.-A. Bachor}
\affiliation{Department of Physics, Faculty of Science, 
The Australian National University, A.C.T. 0200, Australia.}

\author{W. P. Bowen}
\affiliation{Department of Physics, Faculty of Science, 
The Australian National University, A.C.T. 0200, Australia.}

\author{P. K. Lam}
\affiliation{Department of Physics, Faculty of Science, 
The Australian National University, A.C.T. 0200, Australia.}

\author{K. Danzmann}
\affiliation{Max-Planck-Institut f\"ur Gravitationsphysik (Albert-Einstein-Institut),\\
Institut f\"ur Atom- und Molek\"ulphysik, Universit\"at Hannover, 30167 Hannover, Germany}

\email{Roman.Schnabel@aei.mpg.de}

\begin{abstract} 
Quantum noise of the electromagnetic field is one of the limiting noise sources in interferometric gravitational wave detectors. Shifting the spectrum of squeezed vacuum states downwards into the acoustic band of gravitational wave detectors is therefore of challenging demand to quantum optics experiments.
We demonstrate a system that produces nonclassical continuous variable states of light that are squeezed at sideband frequencies below 100~kHz. A single optical parametric
amplifier (OPA) is used in an optical noise cancellation scheme
providing squeezed vacuum states with coherent bright phase
modulation sidebands at higher frequencies. The system has been stably
locked for half an hour limited by thermal stability of our
laboratory. 
\end{abstract}

\maketitle   

Currently, an international array of first-generation, kilometer-scale laser interferometric gravitational-wave detectors, consisting of GEO\,600~\cite{geo02}, LIGO~\cite{LIGO}, TAMA\,300~\cite{TAMA} and VIRGO~\cite{VIRGO}, targeted at gravitational-waves (GW) in the acoustic band from 10~Hz  to 10~kHz, is going into operation. 
These detectors are all Michelson interferometers. Intense laser light is injected from the bright port, whereas the output port is locked to a dark fringe. The anti-symmetric mode of arm-length oscillations (e.g.\ excited by a gravitational wave) yields a sideband modulation field in the anti-symmetric (optical) mode which is detected at the dark output port.
In general, several technical noise sources, and more fundamentally, thermal noise 
and quantum noise (radiation pressure and shot-noise \cite{KLMTV01}, \cite{HCCFVDS03}) contribute to the signal noise floor. 
Recently it has been shown that thermal noise and radiation pressure noise might be sensed by additional short high-finesse cavities and  subsequently reduced \cite{CHP99}, \cite{CHP03}. Shot-noise on the other hand can be reduced by squeezed vacuum states of light injected into the dark port of the interferometer (\cite{SHSD04} and references therein). Experimental progress has been reported in \cite{KSMBL02} where the shot-noise of a power-recycled table-top interferometer has been reduced by squeezed states at about 5~MHz. 
GW interferometers require squeezed states at frequencies of the acoustic detection band. Such states have not been demonstrated so far since 
laser sources are classically noisy in the kHz-regime and below. Noise cancellation schemes have been proposed to enhance squeezing
utilizing Kerr non-linearity in fibers \cite{Shirasaki90}, laser diodes  \cite{LHY91} and
second harmonic generation \cite{Ralph95}.  In a recent experiment, continuous wave squeezing
at frequencies as low as 220 kHz was demonstrated \cite{BSTBL02} utilizing 
two squeezed beams from two independent optical parametric amplifiers (OPAs).

%
%
\begin{figure}[ht]
  \begin{center}
  \includegraphics[width=7.5cm]{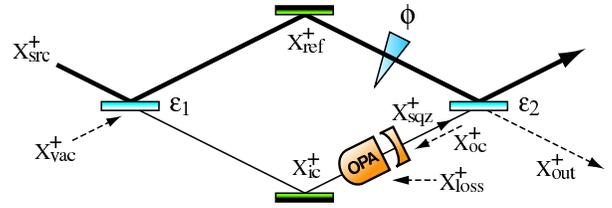}
  \end{center}
  \caption{Schematic of our experiment to produce squeezed vacuum states of light below 100 kHz utilizing a single OPA. 
  }
  \label{TheoryFig}
\end{figure}

In this paper we report the observation of a squeezed vacuum state at a sideband frequency of 80~kHz utilizing a single dim squeezed beam and a classically correlated bright coherent laser beam in a Mach-Zehnder configuration. Broadband vacuum squeezing from 130~kHz up to several MHz was also measured. 
Consider a Mach-Zehnder interferometer with independently adjustable beamsplitter reflectivities $(\varepsilon_{1},\varepsilon_{2})$ and an OPA in one arm of the interferometer (Fig.~\ref{TheoryFig}).  
We theoretically and experimentally show that this configuration can be operated as a common mode noise cancellation experiment providing a squeezed vacuum with phase modulation sidebands at the interferometer output.
First let us derive the expression for the amplitude quadrature operator in frequency space $\hat X^{+}_{{\rm out}}$. We consider amplitude noise from our laser source $(\hat X^{+}_{{\rm src}})$ and vacuum fluctuations entering lossy ports of our set-up $(\hat X^{+}_{{\rm vac}}$, $\hat X^{+}_{{\rm loss}}$, $\hat X^{+}_{{\rm oc}})$. 
A bright coherent state represented by the quadrature operator $\hat{X}^{+}_{\mathrm{src}}$ encounters the first beamsplitter $\varepsilon_{1}$ that couples in the vacuum state entering through the unused input port, thereby producing two fields $\hat{X}^{+}_{\mathrm{ic}}\!=\! \sqrt{\varepsilon_{1}}\hat{X}^{+}_{\mathrm{vac}}\!+\!\sqrt{1\!-\!\varepsilon_{1}}\hat{X}^{+}_{\mathrm{src}}$ and $\hat{X}^{+}_{\mathrm{ref}}\!=\!\sqrt{1\!-\!\varepsilon_{1}}\hat{X}^{+}_{\mathrm{vac}}\!-\!\sqrt{\varepsilon_{1}}\hat{X}^{+}_{\mathrm{src}}$. The field $\hat{X}^{+}_{\mathrm{ic}}$ is used to seed a single OPA consisting of a $\chi^{(2)}$ non-linear medium inside a highly over-coupled optical resonator. In our scheme the OPA acts as a de-amplifier resulting in a dim amplitude squeezed transmitted output field $\hat X^{+}_{{\rm sqz}}$. In this configuration the overall non-linearity $g$ of the OPA resonator, which is dependent on the crystal non-linearity, second harmonic pump power and mode matching of fundamental and pump beams, becomes real and negative.  
We follow the treatment of optical parametric oscillation and amplification which uses the linearized formalism of quantum mechanics and the mean field approximation as given in \cite{Lam99} yielding the OPA transfer function 
\begin{eqnarray}
\hat{X}^{+}_{\mathrm{sqz}}&\!=\!&\Big\{\sqrt{4\kappa_{\mathrm{ic}}\kappa_{\mathrm{oc}}}\hat{X}^{+}_{\mathrm{ic}}\!+\!\sqrt{4\kappa_{\mathrm{loss}}\kappa_{\mathrm{oc}}}\hat{X}^{+}_{\mathrm{loss}}\nonumber\\
&&+(2\kappa_{\mathrm{oc}}\!-\!i\Omega\!-\!\kappa\!+ \!g)\hat{X}^{+}_{\mathrm{oc}}\Big\}/(i\Omega\!+\!\kappa\!-\! g)
\end{eqnarray}
where $\kappa_{\mathrm{ic}}$, $\kappa_{\mathrm{oc}}$, $\kappa_{\mathrm{loss}}$ are the input, output and loss coupling rates respectively for the OPA resonator with total decay rate $\kappa\!=\!\kappa_{\mathrm{ic}}\!+\!\kappa_{\mathrm{oc}}\!+\!\kappa_{\mathrm{loss}}$. The sideband frequency of detection is set by $\Omega/2\pi$. The reference beam $\hat{X}^{+}_{\mathrm{ref}}$ is given a phase shift $\phi$ before being interfered with the squeezed beam $\hat{X}^{+}_{\mathrm{sqz}}$ on the second beamsplitter $\varepsilon_{2}$ to give $\hat{X}^{+}_{\mathrm{out}}\!=\!\sqrt{\varepsilon_{2}}\hat{X}^{+}_{\mathrm{sqz}}\!+\!e^{-i\phi}\sqrt{1\!-\!\varepsilon_{2}}\hat{X}^{+}_{\mathrm{ref}}$ on the chosen output port. Fully expanding this expression and collecting terms it becomes clear that
\begin{eqnarray}
\hat{X}^{+}_{\mathrm{out}}&\!=\!&\biggl\{
\hat{X}^{+}_{\mathrm{src}}\Bigl[\sqrt{(1\!-\!\varepsilon_{1})\varepsilon_{2}}\sqrt{4\kappa_{\mathrm{ic}}\kappa_{\mathrm{oc}}}\nonumber\\
&&-\sqrt{\varepsilon_{1}(1\!-\!\varepsilon_{2})}(i\Omega\!+\!\kappa\!-\! g)e^{-i \phi}\Big]\nonumber\\
&&+\hat{X}^{+}_{\mathrm{vac}}\Bigl[\sqrt{\varepsilon_{1}\varepsilon_{2}}\sqrt{4\kappa_{\mathrm{ic}}\kappa_{\mathrm{oc}}}\nonumber\\
&&+\sqrt{(1\!-\!\varepsilon_{1})(1\!-\!\varepsilon_{2}\!)}(i\Omega\!+\!\kappa\!-\! g)e^{-i \phi}\Big]\nonumber\\
&&+\hat{X}^{+}_{\mathrm{oc}}\Bigl[\sqrt{\varepsilon_{2}}(2\kappa_{\mathrm{oc}}\!-\!i\Omega\!-\!\kappa\!+\! g)\Big]\nonumber\\
&&+\hat{X}^{+}_{\mathrm{loss}}\Bigl[\sqrt{\varepsilon_{2}}\sqrt{4\kappa_{\mathrm{loss}}\kappa_{\mathrm{oc}}}\Big]\bigg\}/(i\Omega\!+\!\kappa\!-\! g)
\end{eqnarray}
The value of $\phi$ is set to zero to ensure that complete cancellation of the coherent amplitude at the chosen output port is possible. The sideband detection frequencies are assumed to be within the linewidth of the OPA resonator such that the approximation $\Omega\ll\kappa$ is valid. 
From the above equation the noise spectrum of the output field as measured by a homodyne detector may be calculated using $V^{+}_{\mathrm{out}}\!=\!\langle (\hat{X}^{+}_{\mathrm{out}})^{2} \rangle \!- \!\langle \hat{X}^{+}_{\mathrm{out}} \rangle^{2}$.
The input vacuum states are assumed to be uncorrelated amongst themselves and with variances set to $V^{+}_{\mathrm{vac}}\!=\!V^{+}_{\mathrm{oc}}\!=\!V^{+}_{\mathrm{loss}}\!=\!1$. The laser source contribution $(\hat{X}^{+}_{\mathrm{src}})$ to the output field may be completely removed provided that the following condition, consisting of a relation between both beamsplitter reflectivities and OPA resonator properties, is satisfied:
\begin{equation}
\varepsilon_{1}^{+}=1\!-\!\left[1\!+\!\frac{\varepsilon_{2}}{(1\!-\!\varepsilon_{2})}\frac{4\kappa_{\mathrm{ic}}\kappa_{\mathrm{oc}}/\kappa^{2}}{(1\!-\!g/\kappa)^{2}} \right]^{-1}
\end{equation}
This condition can be interpreted as $\varepsilon_{1}$ compensating for the classical OPA gain in order to achieve perfect interference visibility. With $\varepsilon_{1}$ suitably adjusted, the output noise spectrum becomes a squeezed vacuum of variance $V^{+}_{\mathrm{sqzvac}}$. Here the superscript $^+$ describes the fact that a phase reference might still be given by a modulation field at higher frequencies.
\begin{equation}
V^{+}_{\mathrm{out}}(\varepsilon_{1}=\varepsilon_{1}^{+})=V^{+}_{\mathrm{sqzvac}}
=1\!+\!\varepsilon_{2}\frac{4\kappa_{\mathrm{oc}} g}{(\kappa-g)^{2}}
\label{Vout}
\end{equation}
It can be seen that $\varepsilon_2$ should be close to one to keep losses on the squeezing as small as possible. In the experiment described below we chose $1 - \varepsilon_2 = 1 \%$, which is already a small value in comparison with typical losses in current squeezing experiments.
Note that the noise variance in Eq.~(\ref{Vout}) is identical to that of a single OPA seeded with a quantum noise limited input and detected with $1 - \varepsilon_{2}$ intensity loss. 
However, here a modulation performed in only one interferometer arm will endow the squeezed vacuum with bright modulation sidebands outside the squeezing band of interest, thereby facilitating phase locking of any down-stream applications.

\begin{figure}[ht]
  \begin{center}
  \includegraphics[width=8.5cm]{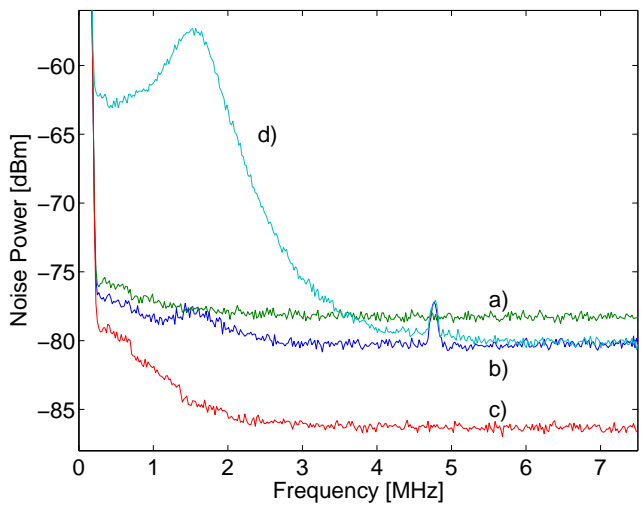}
  \end{center}
  \vspace{-5mm}
  \caption{Measured noise power spectra at sideband frequencies $\Omega/2\pi$. The distances between the shot-noise curve (a) and curves (b) and (d), respectively,  represent the directly observed squeezing. The distance between curves (b) and (d) represents the optical cancellation of technical noise achieved. (c) represents the detector dark noise.}
  \label{OverviewSpectrum}
\end{figure}

\begin{figure}[ht]
  \begin{center}
  \includegraphics[width=8.5cm]{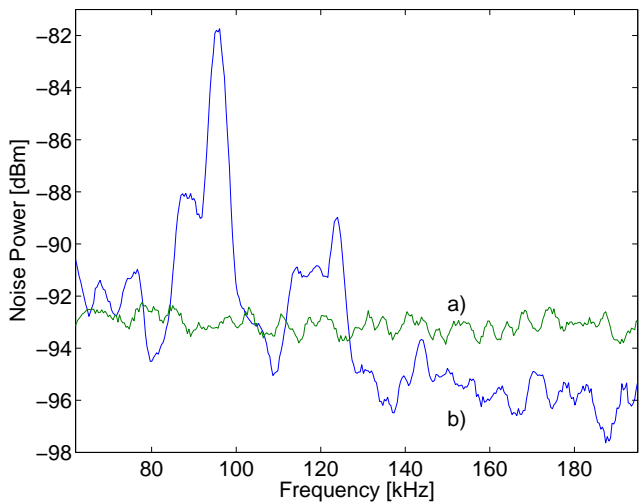}
  \end{center}
  \vspace{-2mm}
  \caption{Measured noise power spectra. Squeezing was observed where curve (b) was below the shot-noise curve (a). 
}
  \label{80kHzSpectrum}
\end{figure}

Figs. \ref{OverviewSpectrum} and \ref{80kHzSpectrum} show experimental results from the single OPA noise cancellation scheme according to Fig.~\ref{TheoryFig}. The OPA was constructed from a non-critically phasematched MgO:LiNbO$_{3}$ crystal inside a hemilithic resonator of input and output power reflectivities of 0.9997 and 0.95, respectively.
The laser source of the experiment was a monolithic non-planar Nd:YAG ring laser of up to 2~Watt single mode output power at 1064 nm. Intensity noise below 2~MHz was reduced by a servo loop acting on the pump diode current. 
The OPA was seeded with a coherent beam of 30~mW at the fundamental wavelength and pumped with 300~mW of the second harmonic (532~nm). 
The second harmonic beam was generated in a cavity which was similar to the OPA cavity. Conversion efficiency of 65\% was achieved.
The phase difference of fundamental and second harmonic waves inside the OPA were stably locked to deamplification generating a dim amplitude quadrature squeezed beam of about 200~$\mu$W at 1064~nm. The noise power spectrum was measured in a homodyne detector constructed from two optically and electronically matched ETX~500 photodiodes. 
Amplitude quadrature squeezing was observed at sideband frequencies from 4 MHz up to the 29~MHz OPA cavity linewidth. Curve (d) in Fig.~\ref{OverviewSpectrum}  shows the noise power spectrum of the amplitude quadrature stably locked to the homodyning local oscillator. 
The locking loop error signal was extracted from 20~MHz phase-modulation sidebands generated by application of a RF electric field along the optical axis of the non-linear OPA crystal.
The shot-noise reference given by curve (a) was measured by blocking the squeezed beam before the homodyne detector. The apparent increase in shot-noise level at lower frequencies is due to higher homodyne detector dark noise, cf. curve (c).
In a second step the dim squeezed beam was overlapped on a beamsplitter of reflectivity $\varepsilon_2 = 99 \%$ with a coherent beam from the same laser source. The intensities and the relative phase of both inputs were adjusted to provide a dark output of less than 6~$\mu$W. The amplitude noise spectrum of the dark port is shown in Fig.~\ref{OverviewSpectrum} curve (b). The spectra in Fig.~\ref{OverviewSpectrum}  were recorded on a spectrum analyzer with resolution bandwidth (RBW) set to 100 kHz and video bandwidth (VBW) set to 100~Hz. In Fig.~\ref{80kHzSpectrum} the RBW was reduced to 3 kHz and the VBW to 10~Hz. 
Here, measured shot-noise (a) and squeezed vacuum noise (b) are shown after Gaussian weighted averaging within the RBW. The detector dark noise was at least 2.5 dB below the squeezed trace before being subtracted.   
Squeezing at 80~kHz and broadband squeezing from 130~kHz upwards were observed. The squeezed beam still carried the phase-modulation sidebands at 20~MHz and was stably locked to the homodyning local oscillator phase.
%
%
%
%
%

%
%

Within the assumptions of the presented theory, perfect cancellation of technical laser source noise is possible, provided that anti-correlated noise contributions are kept to zero, and matching of the coherent amplitudes and the spatial modes at the combining beam splitter are perfect. A flat spectrum of a constant level of squeezing inside the cavity linewidth is then expected. 
In our experiment, residual classical noise at some frequencies still limited the observation of squeezing to the lowest frequency of 80~kHz. 
Our results, however, were not limited by the strength of optical noise cancellation. Classical noise was suppressed by 25~dB at the 1.5~MHz laser relaxation oscillation (Fig.~\ref{OverviewSpectrum}). 
The strength of the optical noise cancellation was limited by 94.4\% visibility at the 99/1 combining beamsplitter. This limitation also led to the residual power of 6~$\mu$W at the dark output port. Classical noise from the homodyne local oscillator was electronically suppressed by $60$~dB and was not significant for our measurements.
Our results were limited by anti-correlated classical noise, possibly arising from acoustic noise coupling into the optical scheme, locking noise of the OPA, electronic pickup of stray RF fields in the electrodes applied to the OPA crystal, or even noise coupled into the system via the second harmonic pump. The signal at 4.8~MHz was identified to be the beat of two modulation frequencies and indeed picked up by the OPA.
In both figures the lower boundary of noise power was set by the squeezing achieved in the OPA and subsequent losses. Photodiode efficiencies of $(92\pm 2)\%$, homodyne detector visibility of $0.975 \pm 0.003$, propagation losses of $(5 \pm 0.5) \%$ and OPA escape efficiency of $(88 \pm 2)\%$ give an overall loss of 27 \%. 

%
%
In conclusion, we have reported the observation of squeezing at low sideband frequencies down to 80~kHz from a single OPA. The wavelength of the carrier laser field was 1064~nm which is compatible with current GW detectors. In total, just 600~mW laser power was necessary to generate the squeezed states. We note that power requirements linearly scale up with the number of OPAs employed in the scheme.
One goal of further investigations will be the reduction of losses and the effect of green-induced infrared absorption \cite{FKAF01}. Both will enable higher green pump power and therefore increased squeezing strength.
Residual classical noise contributions at low frequencies will also be identified in further investigations that aim to reach the acoustic band of gravitational wave detectors.

This work has been supported by the Deutsche Forschungsgemeinschaft and is part of Sonderforschungsbereich 407.

\end{document}